\documentclass{article}
\usepackage{hyperref}
\usepackage{amsfonts,amssymb}
\usepackage{amsthm}

\newcommand{\rr}{\mathbb{R}}   
\newcommand{\cc}{\mathbb{C}}   
\newcommand{\nn}{\mathbb{N}}   

\newtheorem{theorem}{Theorem }[section]

\newtheorem{lemma}[theorem]{Lemma}
\newtheorem{definition}[theorem]{Definition}

\newtheorem{remark}[theorem]{Remark}

\def\Tr{\rm Tr \; }

\def\cA{{\cal A}}

\def\cH{{\cal H}}

\def\cP{{\cal P}}
\def\cM{{\cal M}}

\def\cL{{\cal L}}
\def\cN{{\cal N}}

\def\e{{\rm e}}

\newcommand{\beq}{\begin{equation}}
\newcommand{\eeq}{\end{equation}}

\def\one{\bbbone}
\def\bbbone{{\mathchoice {\rm 1\mskip-4mu l} {\rm 1\mskip-4mu l}
{\rm 1\mskip-4.5mu l} {\rm 1\mskip-5mu l}}}
\def\bbbc{{\mathchoice {\setbox0=\hbox{$\displaystyle\rm C$}\hbox{\hbox
to0pt{\kern0.4\wd0\vrule height0.9\ht0\hss}\box0}}
{\setbox0=\hbox{$\textstyle\rm C$}\hbox{\hbox
to0pt{\kern0.4\wd0\vrule height0.9\ht0\hss}\box0}}
{\setbox0=\hbox{$\scriptstyle\rm C$}\hbox{\hbox
to0pt{\kern0.4\wd0\vrule height0.9\ht0\hss}\box0}}
{\setbox0=\hbox{$\scriptscriptstyle\rm C$}\hbox{\hbox
to0pt{\kern0.4\wd0\vrule height0.9\ht0\hss}\box0}}}}
\def\bbbe{{\mathchoice {\setbox0=\hbox{\smalletextfont e}\hbox{\raise
0.1\ht0\hbox to0pt{\kern0.4\wd0\vrule width0.3pt
height0.7\ht0\hss}\box0}}
{\setbox0=\hbox{\smalletextfont e}\hbox{\raise
0.1\ht0\hbox to0pt{\kern0.4\wd0\vrule width0.3pt
height0.7\ht0\hss}\box0}}
{\setbox0=\hbox{\smallescriptfont e}\hbox{\raise
0.1\ht0\hbox to0pt{\kern0.5\wd0\vrule width0.2pt
height0.7\ht0\hss}\box0}}
{\setbox0=\hbox{\smallescriptscriptfont e}\hbox{\raise
0.1\ht0\hbox to0pt{\kern0.4\wd0\vrule width0.2pt
height0.7\ht0\hss}\box0}}}}

\begin{document}

\title{{The H\"older Inequality for  KMS States}}

\author{Christian D.\ J\"akel\protect\footnote{christian.jaekel@mac.com}
  \and Florian Robl\protect\footnote{florian.robl@gmail.com} \and
School of  Mathematics, Cardiff University, UK}


\maketitle
\begin{abstract}
We prove a H\"older inequality for  KMS States, which generalises a well-known
trace-inequality. Our results are based on 
the theory of non-commutative $L_p$-spaces. 
\end{abstract}



\section{Introduction}

Trace inequalities have played a key role both in mathematics and quantum
statistical mechanics \cite{Ru,Li,Si}. In recent years numerous trace inequalities have been generalised to  $\sigma$-finite von 
Neumann algebras, for example the Golden-Thompson  and Peierls-Bogolubov inequalities \cite{A2}.
In this short note, we generalize the  H\"older trace-inequality. The latter has been used, for example, by Ruelle to construct interacting
Gibbs states~\cite{Rue1}\cite{Rue2} in a box and then control their thermodynamic limit. 
While trace inequalities are useful for quantum systems constrained to a finite volume,
there are good reasons to abandon the boxes and study  quantum statistical systems  
directly in infinite volume. As the generator of the time evolution will no longer have discrete spectrum, 
trace inequalities can not be applied. Thus the H\"older trace-inequality  has to be replaced by the generalised inequality presented in Section~\ref{main-section}. 
It was pointed out by Fr\"ohlich \cite{Fr1} that
the H\"older inequality given in Section~\ref{main-section} 
also plays a crucial role in the context of thermal quantum field
theory. 

The paper is organised as follows. In Section 2 we recall some basic notions of
Tomita-Takesaki theory and state the main result. Section 3
contains an introduction to non-commutative $L_p$-spaces. Section 4 provides the proof of the
main theorem.

\section{The Main Result}
\label{main-section}

In quantum statistical mechanics, thermal equilibrium states are characterised by the {\em KMS condition} \cite{HHW}, which is (a) a generalisation of the Gibbs condition to systems in infinite volume; (b) formulated in terms of analyticity properties of the correlation functions; and (c) can be derived from first principles, like {\em passivity}~\cite{PW} or 
{\em stability} under small adiabatic perturbations of the dynamics~\cite{HKT-P}.

\begin{definition} 
\label{def1}
Let $\cA$ be a $C^{*}$-algebra
and $\{ \tau_t \}_{t \in \rr}$ be a strongly continuous group of $*$-automorphisms of $\cA$.
A normalised positive linear functional~$\omega_\beta$ over $\cA$ is called a $(\tau,\beta)$-KMS state for the inverse temperature
$\beta >0$,
if for all $A, B \in \cA$ there exists a function $F_{A, B}$,  which is continuous and bounded in the strip $0 \le \Im z \le \beta$
and analytic  in the open strip $0 < \Im z < \beta$, with boundary values given by 
\beq
\label{5}
F_{A, B} (t) = \omega_\beta(A \tau_t (B)) 
\eeq
and $F_{A, B} (t+i \beta) = \omega_\beta( \tau_t (B) A) $ for all $t \in \rr$. 
\end{definition}

The KMS-condition
implies that $\omega_{\beta}$ is invariant under $\tau$ and therefore the latter can be 
unitarily implemented in the GNS representation $(\pi,\cH,\Omega)$ associated to 
the pair $(\cA, \omega_{\beta})$. Weak continuity of $\tau$
ensures the existence of a generator $L$, called the {\em Liouvillean}, 
such that $\pi (\tau_t (A)) \Omega=\e^{-it L} \pi (A) \Omega$ and
$L\Omega=0$. 

As the vector $\Omega$ is  cyclic and separating for the von Neumann
algebra $\cM \doteq \pi(\cA)''$, the algebraic operations on $\cM$ define maps 
on the dense set $\cM \Omega \subset\cH$. Tomita's idea to study the $*$-operation on $\cM$
turned out to be especially fruitful. It leads to an anti-linear operator~$S_\circ$,
	\[ 
	S_\circ  \colon A \Omega \mapsto  A^* \Omega \; , \qquad A \in \cM \; , 
	\]
which is closable, and thus allows a polar decomposition for the closure $S= J \Delta^{1/2}$. 
The anti-linear involution $J$ is called the {\em modular conjugation} and the positive albeit
in general unbounded operator $\Delta$ is called the 
{\em modular operator}. The modular conjugation~$J$ satisfies $J^{*}=J$ and $J^2=\one$,
and induces a $*$-anti-isomorphism $j \colon 
A \mapsto JA^*J$ between the algebra $\cM$ and its commutant $\cM'$
(Tomita's theorem). 

More generally, an arbitrary normal faithful state over a von Neumann algebra~$\cM$ is a $(\sigma, -1)$-KMS state 
with respect to the  modular automophisms $\sigma$ given by $A \mapsto \Delta^{is}A \Delta^{-is}$, $A \in \cM$, $s \in \rr$,   
at temperature $\beta=-1$ (see, e.g., \cite{BR}). 
To be precise, the strong continuity assumption, which is part of Definition \ref{def1}, holds on the restricted $C^*$-dynamical 
system \cite[Proposition 1.18]{Sak} associated to the $W^*$-dynamical system~$(\cM, \sigma)$. Uniqueness of the modular 
automorphism ensures that  $\Delta^{1/2}=\e^{-\beta L /2}$.

The {\it standard positive cone} $\cP^\sharp \subset \cH$ is defined as
\[ \cP^\sharp \doteq Ê\overline { \{ J A J A \Omega : A \in \cM  \} } 
= Ê\overline { \{ \Delta^{1/4} A \Omega  :  A \in \cM^+ \} } ,\]
where the bar denotes norm closure \cite{A}. Consequently, a KMS state on a $C^*$-dynamical system $(\cA, \tau)$ gives rise to 
a von Neumann algebra in {\it standard form}, namely a quadruple $(\cH, \cM, J, \cP^\sharp)$, where 
$\cH$ is a Hilbert space, $\cM $ is a von Neumann algebra, 
$J$ is an anti-unitary involution on $\cH$
and $\cP^\sharp$ is a self-dual cone in $\cH$ such that: 
\begin{itemize}
\item [(i)] $J\cM J = \cM '$; 
\item [(ii)] $JAJ = A^*$ for $A$ in the center of $\cM $; 
\item [(iii)] $J \Psi = \Psi$ for $\Psi \in \cP^\sharp$; 
\item [(iv)] $A JA \cP^\sharp \subset \cP^\sharp$ for $A \in \cM $. 
\end{itemize}  

The vector state induced by $\Omega$ extends the KMS state  $\omega_{\beta}$ 
from $\cA$ to  $\cM$, and we denote this state by the same letter. Now set, for $p \in \nn$ 
and $A \in \cM^+$, 
\begin{equation}
\label{eq1}
|\kern -.5mm |\kern -.5mm | \, A  |\kern -.5mm |\kern -.5mm |_{p }  
\doteq   \omega_\beta \bigl( \underbrace{\e^{it L/p } A \cdots  \e^{it L/p }A }_{p-times} \bigr)_{\upharpoonright t=i \beta}^{1/p} \; .
\end{equation}
The subscript indicates the analytic continuation of the map
$t \mapsto  F(t) \doteq \omega_\beta 
\bigl( \e^{it L/p } A \cdots  \e^{it L/p }A  \bigr)$ 
to $F( i \beta)$. To simplify the notation we will denote $F(i \beta)$ by  
$ \omega_\beta \bigl(  \e^{-\beta L/p } A \cdots  \e^{-\beta L/p }A \bigr)$.

\begin{theorem}[H\"older inequality]
\label{hoelder} Consider a $(\tau,\beta)$-KMS state $\omega_\beta$ over a $C^*$-dynamical system $(\cA, \tau)$.
Let $(z_1, \ldots, z_n)\in \cc^n$ be such, that $0 \le \Re z_j$,
$\sum_{j=1}^n \Re z_j \le 1 $, and
let $p_j$ be the smallest, positive even integer such that
$\frac{1}{p_j}\le \min\{\Re z_{j+1},\Re z_j\}$, with $z_{n+1}=z_n$ and $z_0=z_1$. Then 
\begin{eqnarray}
\label{crucialbound}
\Bigl| \omega_\beta \bigl( A_n  \e^{- z_n \beta L } \cdots A_1\e^{-z_1 \beta L}
A_0 \bigr) \Bigr| \le |\kern -.5mm |\kern -.5mm | \, A_0  |\kern -.5mm |\kern -.5mm |_{p_0} 
\cdots 
|\kern -.5mm |\kern -.5mm | \, A_n  |\kern -.5mm |\kern -.5mm |_{p_n}  
\end{eqnarray}
for all $A_0, \ldots, A_n \in \cM^+$.
\end{theorem}

\paragraph{\bf Remarks} 
\begin{itemize}
\item [(i)] Although the multi-boundary Poisson  kernels \cite[Lemma 4.4.8]{Sak} for the domain $ I^{(n)}$ can be computed explicitly (the computation can be traced back to Widder \cite{Wi}), it seems unlikely that the H\"older inequality 
(\ref{crucialbound}) can be derived using only methods of complex analysis 
(unless $n= 2$).
\item [(ii)] Let $\cM_0$ denote a weakly dense sub-algebra  of analytic\footnote{An element 
$A \in \cM$ is called analytic for $\tau_t$ if there exists a strip
$	I_\lambda = \{ z \in \cc  :  |\Im z | < \lambda \}
$ in $\cc$, and a function $f \colon I_\lambda \mapsto \cM$, such that 
(i) $f(t) = \tau_t (A)$ for $t \in \rr$, and
(ii) $z \mapsto \phi (f(z))$ is analytic for all $\phi \in \cM_*$. } elements
in~$\cM$. It follows that,  for $p \in \nn$ and $A \in \cM_0^+$,
\begin{equation}
\label{lp}
|\kern -.5mm |\kern -.5mm | \, A  |\kern -.5mm |\kern -.5mm |_{p } = \omega_\beta \bigl( \tau_{i \beta/2p } (A) \cdots  \tau_{i (2p-1)\beta/2p }(A) \tau_{i\beta} (A) \bigr)^{1/p} \; .
\end{equation}
Thus Theorem \ref{hoelder} is a generalisation of the H\"older
inequality for Gibbs states, as stated, for example, in \cite{MZa,MZb}.
\end{itemize}

Two more aspects of Theorem \ref{hoelder} are notable. Firstly, it estimates a
non-commutative expression in terms of essentially commutative ones, which can
be evaluated using spectral theory, and secondly, the bounds are uniform in~$\Im
z_j$, $j = 1, \ldots, n$. The proof of Theorem \ref{hoelder} relies on the theory of
non-commutative $L^p$-spaces, but the appeal of the theorem may well
be that knowledge of non-commutative integration theory is not required in order {\em to apply} the inequality.

In quantum statistical mechanics the uniformity in imaginary time is useful
for establishing the existence of real time Greens functions from the
Schwinger functions. Beyond quantum statistical mechanics, inequality (\ref{crucialbound}) is also useful in constructive quantum field theory. In
\cite{Fr1} Fr\"ohlich argued that the H\"older inequality will guarantee the
existence of thermal Wightman functions for a certain class of models. A complete proof of this claim is given in
\cite{JR}. Additionally, in a forthcoming work by M.~Rouleux and the first author, the H\"older inequality is used to show that 
the Wightman distributions of the $P(\phi)_2$ model on the de Sitter space satisfy a micro-local spectrum condition.

\section{Non-commutative $L_p$-spaces}

Normal states over a von Neumann algebras provide a non-commutative extension of classical integration theory,
i.e., commutative $L^p$-spaces, and one  recovers the latter in case the algebra is abelian \cite{Mu}. 
Among the many approaches to non-commutative $L_p$-spaces \cite{Di,H,Hi,Ko,N,Se,Te},
Araki and Masuda's  approach~\cite{AM} is best suited for our purposes. We start with a short 
introduction to relative modular operators for
weights. A more elaborate discussion of relative
modular operators can be found in \cite{Tak}.

\subsection{Relative Modular Operators}
Consider a general ($\sigma$-finite) von Neumann algebra $\cM$ and let 
$\phi$ be a normal semi-finite weight on $\cM$. The semi-cyclic
representation\footnote{The semi-cyclic representation is a generalisation of
  the GNS representation to weights.} \cite{Tak} makes it possible to define an anti-linear operator $S_{\phi, \Omega}$ by
	\[ S_{\phi, \Omega} A \Omega \doteq\xi_\phi (A^*) \; , \qquad A \in  \cN_\phi^* \; ,  \]
where $ \cN_\phi \doteq \{  A \in \cM  :  \phi (A^*A) < \infty \}  $, and $\xi_\phi (A)$ is the semi-cyclic representation of $A \in \cN_\phi$ in 
\[
\cH_\phi \doteq \overline{ \cN_\phi / {\rm ker} \, \phi } \; .
\]
$S_{\phi,\Omega}$  is closable and the closure $\overline{S_{\phi, \Omega}}$ has a polar
decomposition $\overline{S_{\phi, \Omega}} \doteq  J_{\phi,\Omega}\Delta_{\phi,  \Omega}^{1/2}$. It is noteworthy that
	\[	\Delta_{\phi, \Omega} = S_{\phi, \Omega}^* \overline{S_{\phi, \Omega}} \; ,	\]
is a positive, in general unbounded, operator on the {\em original} Hilbert space $\cH$.
If $\phi$ is a vector state associated to $\xi \in \cH$ such that $\phi(x)=(\xi,x\xi)$, then $\xi_\phi (A) = A \xi$ and we denote 
$\Delta_{\phi, \Omega}$ by $\Delta_{\xi, \Omega}$ and $J_{\phi,\Omega}$ by $J_{\xi,\Omega}$. In order to keep the notation simple, 
$\e^{-\beta L/ 2}$ will from
now on be written as $\Delta^{1/2}\equiv \Delta_{\Omega, \Omega}^{1/2}$.

A key role in the proof of Theorem \ref{hoelder}  will be played by the following
estimate: 
define, for any $\alpha>0$, a set 
\begin{equation} \label{Ibeta}
I^{(n)}_\alpha \doteq \{ (z_1, \ldots, z_n) \in \cc^n  : 
\sum_{j=1}^n \Re z_j\le\alpha ,  \, 0 \le \Re z_j  \} \;.
\end{equation}
Let  $z \in I^{(n)} \equiv I_1^{(n)}$ and $z_m', z_m''\in \cc$ be 
such that $\Re z_m' , \Re z_m''>0$, $z_m'+z_m''=z_m$ and
\begin{eqnarray}
\Re z_1 + \ldots \Re z_{m-1} + \Re z_m'' &\le& 1/2 \; ,\label{div1}\\
\Re z_n + \ldots \Re z_{m+1} + \Re z_m' &\le& 1/2 \; .\label{div2}
\end{eqnarray}
Under these conditions, Araki \cite[Lemma A]{AM}  has shown\footnote{Note that, in contrast to \cite {AM}, our  inner product is linear in the second entry.} that  for $\phi_1, \ldots, \phi_n \in \cM_{*}^{+}$ and
$X_0,\ldots,X_n\in \cM$ 
\begin{eqnarray}
\Bigl|(\Delta_{\phi_m,\Omega}^{\bar{z}_m'}X_m^{*}&\Delta&_{\phi_{m+1},\Omega}^{\bar{z}_{m+1}}
\ldots
\Delta_{\phi_n,\Omega}^{\bar{z}_n}X_n^{*}\Omega,
\Delta_{\phi_m,\Omega}^{z_m''}X_{m-1}\Delta_{\phi_{m-1},\Omega}^{z_{j-1}} \ldots
\Delta_{\phi_1,\Omega}^{z_1}X_0\Omega  )\Bigr|\nonumber\\
&\le& \Bigl(\prod_{j=0}^n\|X_j\| \Bigr)
(\Omega , \one \Omega)^{z_0} \Bigl(\prod_{j=1}^n \phi_j(\one)^{\Re z_j} \Bigr) \; ,  \label{arakibound}
\end{eqnarray}
with $z_0=1-\sum_{j=1}^n\Re z_j$.

\begin{remark} Consider the space of $n \times n$-matrices $M_n (\cc) \ni  \xi , \eta$ equipped with the inner product
$( \xi , \eta ) = \Tr \xi^* \eta $  and two positive matrices $0< \nu, \omega \in M_n (\cc)$. 
Moreover, assume that $\Tr \omega =1$. Now apply the H\"older trace inequality \cite{MZa}
\begin{eqnarray}
| \Tr  \omega A B  |\le \| A\|_{\omega,p} \, \| B\|_{\omega,q} \; , \qquad p^{-1} + q^{-1} =1 \; , 
\end{eqnarray}
where $\langle A \rangle_{\omega}\doteq\Tr \omega  A $  and  $\| A
\|_{\omega,p}^p\doteq\Tr (\omega^{1/2p}|A|\omega^{1/2p})^p$,
 to the relative modular operator $\Delta_{\nu, \omega}$, which satisfies  
$\Delta_{\nu,\omega}^{1/p} \xi= \nu^{1/p} \xi \omega^{-1/p}$ for $p \in \nn$. Thus, for $1/p +1/q =1$,  
\begin{eqnarray}
|\langle A_2 \Delta_{\nu_2,\omega}^{1/p} A_1 \Delta_{\nu_1,\omega}^{1/q} A_0
\rangle_{\omega}|
&\le& \Bigl( \prod_{j=0}^2 \| A_j\|_{\infty} \Bigr) \, \|\Delta_{\nu_2,\omega}^{1/p}\|_{\omega,p} \,\|\Delta_{\nu_1,\omega}^{1/q}\|_{\omega,q}
\nonumber \\
\label{hoeldergibbs}
&= &\Bigl( \prod_{j=0}^2 \| A_j\|_{\infty} \Bigr) \, \langle \one
\rangle_{\nu_2}^{1/p} \, \langle \one \rangle_{\nu_1}^{1/q} \; .
\end{eqnarray}
\end{remark}

\subsection{Positive Cones and $L_p$-Spaces for von Neumann Algebras}

Consider a general $(\sigma$-finite) von Neumann algebra $\cM$ in standard
form with cyclic and separating vector $\Omega$.  
For $2 \le p \le \infty$, Araki and Masuda define \cite[Equ.~(1.3), p.~340]{AM}
	\[
	L_p (\cM, \Omega) \doteq \bigl\{ \zeta \in \bigcap_{\xi \in \cH} D \bigl( \Delta_{\xi, \Omega}^{\frac{1}{2} - \frac{1}{p}} 
	\bigr)  :  \| \zeta \|_p < \infty \bigr\},
	\]
where
	\[
	\| \zeta \|_p = \sup_{ \| \xi \|= 1} \| \Delta_{\xi, \Omega}^{\frac{1}{2} - \frac{1}{p}} \zeta \| \; .  
	\]
For $1\le p <2$,  $L_p (\cM, \Omega)$ is defined as the completion of $\cH$ with respect to the norm
\[
\|\zeta\|_p=\inf\{\|\Delta_{\xi,\Omega}^{\frac{1}{2}-\frac{1}{p}} \zeta\| :
\|\xi\|=1, s_{\cM}(\xi)\ge s_{\cM}(\zeta)\} .
\]
Here $s_{\cM}(\xi)$ denotes the smallest projection in $\cM$, which leaves $\xi$
invariant. The cones \cite[Equ.~(1.13)]{AM}
	\[ 
	\cP^\alpha  \doteq \overline{\{ \Delta^{\alpha} A \Omega  :  A \in \cM^+ \}}  \; , \qquad 0 \le \alpha \le 1/2 \; , 
	\]
can be used to define the positive part of $L_p (\cM, \Omega)$ \cite[Equ.~(1.14), p.~341]{AM}:
\begin{eqnarray}
L_p^+ (\cM, \Omega) \doteq L_p  (\cM, \Omega) \cap \cP_\Omega^{1/ (2p)} \; , \qquad 2 \le p \le \infty \; . \label{pospart}
\end{eqnarray}	
%
%
Note that these are not operator spaces. The connection to the operator
algebra~$\cM$ is made through auxilliary spaces $\cL_p (\cM, \Omega)$, which consist of formal expressions $ A = u
\Delta_{\phi, \Omega}^{1/p}$ with $\phi \in \cM_*^+$ and $u$ a partial isometry
satisfying $u^*u =s (\phi)$ (the support projection of $\phi$).  The set  
of formal products
\begin{equation} \label{geschwL}
X_0 \Delta_{\phi_1, \Omega}^{z_1} X_1 \cdots \Delta_{\phi_n, \Omega}^{z_n} X_n,
\end{equation}
is denoted by $\cL_p^{*}(\cM, \Omega)$. Here is $ X_j \in \cM$ ($j= 0, \ldots, n$), $\phi_j \in \cM_*^+$ ($j= 1, \ldots,
n$) and $ \vec z = (z_1, \ldots, z_n) \in I_{1-
  (1/p)}^{(n)}$. On the subset $\cL_{p, 0}^* (\cM, \Omega)\subset\cL_p^{*}(\cM,
\Omega)$,
characterized by the condition $\sum_{j =1}^n \Re z_j = 1 - (1/p)$, it is possible to implement
the star operation. The adjoint of a generic
element (\ref{geschwL}) in $\cL_{p,0}^{*}(\cM,\Omega)$  is
defined  to be
\begin{equation}
X^*_n \Delta_{\phi_n, \Omega}^{\overline{z_n}}  \cdots X_1^* \Delta_{\phi_1, \Omega}^{\overline{z_1}} X_0^*.
\end{equation}
A multiplication can  be defined, using the product in $\cM$ to connect the formal expressions:  
$BC \in \cL_{r, 0}^* (\cM, \Omega)$ for $B \in
\cL_{p, 0}^* (\cM, \Omega)$, $C \in \cL_{q, 0}^* (\cM, \Omega)$ and $r^{-1} =
p^{-1} + q^{-1} - 1$. 

If $r^{-1}= \sum_{j=1}^n (p_j)^{-1} $, 
$ r^{-1} + r^{-1} = 1$, $ \xi_j \in L_{p_j} (\cM, \Omega)$, $X_j \in \cM$
$(j=0, \ldots, n)$, and $\xi_j = u_j \phi_j^{1/p_j}$, 
$(j=1, \ldots, n)$ is the polar decomposition, then the product
\[
\xi = X_0 \xi_1 X_1 \xi_2 \cdots \xi_n X_n \in L_r (\cM, \Omega)  
\qquad (= L_{r'} (\cM, \Omega)^*) 
\]
is defined by 
\[
\langle\xi , \xi' \rangle = 
\omega ( \Delta_{\phi', \Omega}^{1/r'} u'^* X_0 u_1  \Delta_{\phi_1, \Omega}^{1/p_1} X_1 u_2 
\Delta_{\phi_2, \Omega}^{1/p_2} \cdots 
u_n  \Delta_{\phi_n, \Omega}^{1/p_n} X_n ) \in L_r (\cM, \Omega)  
\]
where $\xi' \in L_{r'}(\cM , \Omega)$ and $\xi' = u' {\phi'}^{1/r'}$ is its polar decomposition.

Araki's inequality (\ref{arakibound}) now entails a H\"older inequality: let $\zeta_1 \in L_p(\cM,\Omega)$ and $\zeta_2 \in
L_{p'}(\cM,\Omega)$ for $p^{-1}+p'^{-1}=r^{-1}$, then
\begin{equation}
\label{Lholder}
\|\zeta_1 \zeta_2\|_r\le \|\zeta_1\|_p \, \|\zeta_2\|_{p'}.
\end{equation}
Thus the product $\zeta_1\zeta_2$ is in
$L_r(\cM,\Omega)$ and, as the case $p^{-1}+p'^{-1}=1$ suggests, the
topological dual $L_p(\cM,\Omega)^{*}$ of $L_p(\cM,\Omega)$  is
$L_{p'}(\cM,\Omega)$. 
For $A\in
\cL_p(\cM,\Omega)$ and $B \in \cL_p(\cM,\Omega)^{*}$, the corresponding duality bracket
is given by
\begin{equation}
\langle A ,B \rangle = (A \Omega, B \Omega) \; ,
\end{equation}
if $\Omega$ is in the domain of $A$ and $B$. 
According to \cite[Notation 2.3
(4)]{AM} $A$ and $B$ in
$\cL_p^{*}(\cM,\Omega)$ are said to be equivalent, if  (i)  $1\le p \le 2$ and $A\Omega=B\Omega$; (ii) if
$2 \le p \le \infty$ and  
\beq 
\langle C,A \rangle=\langle C, B \rangle
\eeq
 for all $C$ in
$\cL_p(\cM,\Omega)$.

Another important
property is, that for $1\le p\le \infty$, $x \in \cM$ and $\zeta \in
L_p(\cM,\Omega)$, the following inequality holds:
\begin{equation}
\|x\zeta\|_p\le \|x\| \, \|\zeta\|_p.
\end{equation}
It is evident from the definition of the $L_p$-spaces, that $\cH$ and $L_2(\cM,\Omega)$ are equal. It is
proven in \cite{AM} that $\cM \cong L_{\infty}(\cM,\Omega)$ as well as $\cM_{*} \cong L_1(\cM,\Omega)$.

\section{Proof of the Main Result}

\begin{lemma}
\label{lemma1}
Let $A_1, \ldots, A_n \in \cM^+$. Then there exist unique $\phi_j \in \cM_*^+$
such that for $0 \le p_j^{-1} \le 1/2$
\begin{eqnarray}
\Delta_{\phi_j, \Omega}^{1/p_j} \Omega = \Delta^{1/2p_j} A_j \Omega \;   
\qquad (j=1, \ldots, n)
\end{eqnarray}
and $ \phi_j(\one)^{1/p_j} =  \| \Delta^{1/2p_j} A_j \Omega \|_{p_j} $. 
If also $\sum_{j=1}^n 1/p_j = 1/2$ holds, then  
\begin{eqnarray}
\label{27}
 \Delta_{\phi_n, \Omega}^{1/p_n} \cdots \Delta_{\phi_1, \Omega}^{1/p_1}  \Omega =
\Delta^{1/2p_n} A_n \Delta^{1/2p_n}\cdots \Delta^{1/2p_1} A_1 \Omega \; \in \cH.  
\end{eqnarray}
\end{lemma}

\begin{proof}
Let  $A_1, \ldots, A_n \in \cM^+$ and $0 \le p_j^{-1} \le 1/2$, $j=1, \ldots, n$. Then, by definition $\zeta_j \doteq
\Delta^{1/2p_j} A_j \Omega \in \cP_\Omega^{1/2p_j}$. An application of
inequality (\ref{arakibound}) yields
\begin{eqnarray}
 \| \zeta_j\|_{p_j}^2 &=&  \sup_{ \| \xi \|= 1} \| \Delta_{\xi, \Omega}^{(1/2) - (1/p_j)} \zeta_j \|^2 
 \\
 &=& \sup_{ \| \xi \|= 1}   \left(  \Delta_{\xi, \Omega}^{(1/2) - (1/p_j)} \Delta^{1/2p_j} A_j \Omega , 
 \Delta_{\xi, \Omega}^{(1/2) - (1/p_j)} \Delta^{1/2p_j} A_j \Omega \right)
 \\
 &\le & \sup_{ \| \xi \|= 1}  (\xi , \one \xi)^{1- (2/p_j)} \omega (\one)^{2/p_j} \| A _j\|^2  =  \| A_j \|^2 < \infty\; ,
 \end{eqnarray}
which establishes, that $\zeta_j  \in L_{p_j} (\cM, \Omega)$. Thus, according
to (\ref{pospart}), $\zeta_j \in
L_{p_j}^+ (\cM, \Omega)$. By \cite[Theorem 3  (4), p.~342]{AM} there exists a unique $\phi_j \in \cM_*^+$
such that $\zeta_j = \Delta_{\phi_j, \Omega}^{1/p_j} \Omega$ and $
\phi_j(\one)^{1/p_j} = \| \zeta_j \|_{p_j} = \| \Delta^{1/2p_j} A_j \Omega
\|_{p_j} $.

Thus, by definition \cite[Notation 2.3 (4)]{AM}, $\Delta^{1/2p_j} A_j \Delta^{1/2p_j}
\equiv \Delta_{\phi_j, \Omega}^{1/p_j} $  as elements in  $\cL_{p_j',0}^*(\cM,
\Omega)$, where $p_j^{-1}+p_j'^{-1}=1$. Even though $\Delta_{\phi_j,\Omega}^{1/p_j}$
and $\Delta^{1/2p_j}A\Delta^{1/2p_j}$ may not be equal as operators, Lemma 7.7 (2) in \cite{AM}
ensures, that their composition as elements of the spaces $\cL_p^{*}$ is
well-defined: setting $B_1=
\Delta_{\phi_2, \Omega}^{1/p_2}$, $B_2 = - \Delta^{1/2p_2} A_2 \Delta^{1/2p_2}$ and 
$C_2 = \Delta_{\phi_1, \Omega}^{1/p_1} $, there holds $\sum_{i=1}^2 B_i = 0$ as
elements in $L_{p_2}(\cM, \Omega)$, and therefore, using the lemma cited,
\begin{eqnarray}
\label{36}
\Delta_{\phi_2, \Omega}^{1/p_2} \Delta_{\phi_1, \Omega}^{1/p_1} \Omega \equiv
\Delta^{1/2p_2} A_2 \Delta^{1/2p_2}\Delta_{\phi_1, \Omega}^{1/p_1} \Omega
\end{eqnarray}
as elements in $L_{r_1}(\cM,\Omega)=L_{r_1'} (\cM, \Omega)^*$, where
$r_1^{-1}+r_1'^{-1}=1$, $r_1'^{-1}= {p_1'}^{-1} +
p_2'^{-1}-1$ and $1\le r_1' \le 2$ (in comparison to \cite{AM} indices and primed indices have
swaped places). Note that this means $r_1^{-1}=p_2^{-1}+p_1^{-1}$. Using the
same lemma once more (with the appropriate choices of $C_2$ and $B_3$, $B_4$) gives
\begin{eqnarray}
\label{37}
\Delta^{1/2p_2} A_2\Delta^{1/2p_2} \Delta_{\phi_1, \Omega}^{1/p_1} \Omega \equiv  \Delta^{1/2p_2} A_2\Delta^{1/2p_2} \Delta^{1/2p_1} A_1 \Omega
\end{eqnarray}
as elements in $L_{r_1'} (\cM, \Omega)^*$. Together (\ref{36}) and (\ref{37}) imply
\begin{eqnarray}
\Delta_{\phi_2, \Omega}^{1/p_2} \Delta_{\phi_1, \Omega}^{1/p_1} \Omega \equiv
\Delta^{1/2p_2} A_2 \Delta^{1/2p_2} \Delta^{1/2p_1} A_1 \Omega
\end{eqnarray}
as elements in $L_{r_1'} (\cM, \Omega)^*$. Consequently,
\begin{eqnarray}
\Delta_{\phi_2, \Omega}^{1/p_2} \Delta_{\phi_1, \Omega}^{1/p_1} \equiv
\Delta^{1/2p_2} A_2\Delta^{1/2p_2} \Delta^{1/2p_1} A_1 \Delta^{1/2p_1},
\end{eqnarray}
as elements in $\cL_{r_1',0}^{*}(\cM,\Omega)$. Iteration of this procedure results in
\begin{eqnarray}
 \Delta_{\phi_n, \Omega}^{1/p_n} \cdots \Delta_{\phi_1, \Omega}^{1/p_1}  \Omega \equiv 
\Delta^{1/2p_n} A_n\Delta^{1/2p_n} \cdots \Delta^{1/2p_2} A_2\Delta^{1/2p_2} \Delta^{1/2p_1} A_1   \Omega
\end{eqnarray}
as elements in $L_{2} (\cM, \Omega)^*$, because of $\sum_{j=1}^n 1/p_j = 1/2$. But
since $\cH=\cH^{*}=L_{2} (\cM, \Omega)^*$ the proof is finished.
\end{proof}

\begin{lemma}
\label{lemma2}
Let $p \in \nn$ be even and $A \in \cM^+$. Then there
exists $\phi\in\cM_{*}^{+}$ such that
 \begin{eqnarray}
\label{wichtig}
 \| \Delta^{1/2p} A \Omega \|_p=\phi(\one)^{1/p}=\omega_{\beta} ( A \Delta^{1/p} A
\cdots \Delta^{1/p} A)^{1/p}. 
 \end{eqnarray}
On the r.h.s.~we have used Araki's symbolic notation introduced in the sentence following Equ. (\ref{eq1}).
\end{lemma}

\begin{proof}
As proved in Lemma \ref{lemma1}, there exists $\phi\in\cM_{*}^{+}$, such that
$\| \Delta^{1/2p} A \Omega
\|_p^p=\phi(\one)$, and $\Delta^{1/2p} A\Delta^{1/2p} \equiv \Delta_{\phi, \Omega}^{1/p}$
as elements in $\cL^*_{p,0} (\cM, \Omega)$. Thus, by (\ref{27}) and~(\ref{arakibound}),
\begin{eqnarray}
\omega_{\beta} (  \Delta^{1/2p} A \Delta^{1/2p} \cdots  \Delta^{1/2p} A  \Delta^{1/2p})\nonumber 
&=&( \Delta_{\phi, \Omega}^{1/p}  
\cdots \Delta_{\phi, \Omega}^{1/p}\Omega ,   \Delta_{\phi, \Omega}^{1/p}  
\cdots \Delta_{\phi, \Omega}^{1/p}\Omega)  \nonumber \\
&\le&   \phi(\one)  = \| \Delta^{1/2p} A \Omega \|_p^p \; . 
\end{eqnarray}
Since $\phi\in\cM_{*}^{+}$, there exists~\cite{BR} a vector $\xi\in\cP^{\sharp}$ such
that $\phi(X)=(\xi,X\xi)$ for $X \in \cM$. Using $\xi=J_{\phi,\Omega}\Delta_{\phi,\Omega}^{1/2}\Omega=J_{\xi,\Omega}\Delta_{\xi,\Omega}^{1/2}\Omega$, there holds
\begin{eqnarray*}
\phi(X)=(\xi,X\xi)= (\Delta_{\phi,\Omega}^{1/2} \Omega,J_{\phi,\Omega}^{*}
J_{\phi,\Omega}\Delta_{\phi,\Omega}^{1/2} X^{*}\Omega) \; ,
\end{eqnarray*}
where $J_{\phi,\Omega}^{*} J_{\phi,\Omega}=s_{\cM}(\xi)\, s_{\cM'}(\Omega)$ is a projection
\cite[p. 396]{AM}. Therefore 
\begin{eqnarray*}
\phi(\one)\le(\Delta_{\phi,\Omega}^{1/2} \Omega,
\Delta_{\phi,\Omega}^{1/2}\Omega) 
=\omega_{\beta}( A \Delta^{1/p} A
\cdots \Delta^{1/p} A) \; ,
\end{eqnarray*}
which finishes the proof.
\end{proof}

\bigskip
\noindent
{\em Proof of Theorem \ref{hoelder}.} Assuming the requirements of Theorem \ref{hoelder}, Lemma \ref{lemma1} together 
with inequality (\ref{arakibound}), relation (\ref{wichtig}) and $w_j=z_j-(2p_j)^{-1}-(2p_{j-1})^{-1}$ imply
\begin{eqnarray*}
\Bigl| \omega_{\beta}(A_n\Delta^{z_n}\ldots A_1\Delta^{z_1} A_0)\Bigr|
&=& \Bigl|
\omega_{\beta}(\Delta^{1/2p_n}A_n\Delta^{1/2p_n}\Delta^{w_n}
\ldots \Delta^{1/2p_0} A_0 \Delta^{1/2p_0})\Bigr|\\
&=& \Bigl|
\omega_{\beta}(\Delta_{\phi_n, \Omega}^{1/p_n} \Delta^{w_n} \cdots \Delta_{\phi_1, \Omega}^{1/p_1} \Delta^{w_1}
\Delta_{\phi_0, \Omega}^{1/p_0})\Bigr|\\
&\le& \omega_{\beta}(\one)^{1-\sum_{j=0}^n (p_j)^{-1}} \, \prod_{j=0}^n
\phi_j(\one)^{1/p_j} =\prod_{j=0}^n |||A_j|||_{p_j} \; .
\end{eqnarray*}
Again we have used Araki's symbolic notation introduced in the sentence following Equ. (\ref{eq1}).
\qed

\paragraph{Acknowledgement}

This work was supported by the Leverhulme Trust [RCMT 154].

\end{document}